\documentstyle[prl,multicol,aps,psfig]{revtex}
\newcommand{\bea}{\begin{eqnarray}}
\newcommand{\eea}{\end{eqnarray}}

\begin{document}

\title{Precision Monte Carlo Test of the Hartree-Fock Approximation 
for a trapped Bose Gas}
\author{Markus Holzmann and Werner Krauth$\;^*$}
\address{CNRS-Laboratoire de Physique Statistique de l'Ecole 
Normale Sup{\'e}rieure,
24, rue Lhomond, 75231 Paris Cedex 05, France}
\author{Martin Naraschewski
\footnote{holzmann@physique.ens.fr; 
krauth@physique.ens.fr;  martin@mistral.harvard.edu}}
\address{Jefferson Laboratory, Department of Physics, 
Harvard University, Cambridge MA~02138, USA}
\address{ITAMP, Harvard-Smithsonian Center for Astrophysics, 60 Garden Street,
Cambridge, MA~02138, USA
}
\maketitle

\begin{abstract}
We compare 
the semiclassical Hartree-Fock approximation for a trapped Bose gas to a
direct Path Integral Quantum Monte Carlo simulation. The chosen parameters
correspond to current $^{87}$Rb experiments. We observe corrections
to the mean-field density profile. 
The Path Integral calculation reveals an increase of the 
number of condensed particles, 
which is of the same order as a previously computed 
result for a homogeneous system.
We discuss the experimental observability of
the effect and propose a method to analyze data of {\em in-situ}
experiments.
\end{abstract}

\begin{multicols}{2}
\narrowtext
The experimental realization of Bose-Einstein condensation (BEC)
in dilute atomic vapors \cite{anderson95,davis95,bradley95}
has generated extraordinary  experimental and theoretical interest. 
 From the viewpoint of  quantum many-body physics, the trapped
atomic vapors are peculiar. Well above the critical point,
the gases are extremely dilute, and their description as noninteracting
bosons is very accurate. As the condensation sets in, the trapped
atoms are strongly compressed in real space.  Then, interactions become 
much more important and even the simplest
thermodynamic quantities (spatial distribution, condensate fraction,
etc.) have to be obtained by the appropriate 
quantum many-body technique.  At zero temperature the Bogoliubov
approach of weakly interacting Bose gases is well established
\cite{bogoliubov47}. There, the macroscopic condensate
wave function is given by the Gross-Pitaevskii equation
\cite{gross61}. The effects of noncondensed particles
at finite temperatures can be included {\em via}  the
Hartree-Fock-Bogoliubov equations, which have been solved in the
Popov approximation \cite{hutchinson97} and in various simplified
forms \cite{dalfovo98}.

The equilibrium properties of Bose gases can also be directly
computed by Path Integral Quantum Monte Carlo (QMC) simulation
without any essential approximation \cite{ceperley}.
The QMC approach was used to first show \cite{krauth96} that the
trapped Bose gas with repulsive interactions has a lower critical
temperature $T_c$ and a smaller number $N_0$ of condensed particles
below $T_c$ than the free Bose gas, one of the predictions of
mean-field theories \cite{goldman81,giorgini97}.

Very importantly, the QMC calculation is free of systematic errors:
it gives a somewhat noisy, but otherwise  exact numerical
solution of BEC. For simulations corresponding to dilute atomic
vapors, the QMC calculation can be performed directly for the large
particle numbers ($\sim 10^4 - 10^5$) and the temperatures of
experimental interest.  In this situation, there is little  need
to perform finite-size scaling. QMC provides us thus with
a unique opportunity to check the above-mentioned many-body techniques
directly in an experimental setting, a comparison which we provide
here.

It is the main point of this paper to study the {\em corrections}
of a full many-particle treatment to a mean-field theory.  We
compare the QMC calculations to the semiclassical Hartree-Fock (HF)
approximation, which determines the local density in accordance
with the local trap potential and the mean interaction energy
between the particles.  The HF approximation provides an ideal
reference system, since it corresponds locally to a homogeneous
ideal Bose gas. The corrections to HF should be therefore comparable
to previous calculations in a homogeneous system
\cite{Franck,Stoof,houbiers97}. 

For a system closely corresponding to many experiments with $^{87}$Rb
we find  deviations from the HF density which are concentrated in
the overlap region between the excited atoms and the condensate.
The number of  condensed particles $N_0$ is increased around the
critical temperature by about $5 \%$.  We also discuss a scheme
for a detailed analysis of experimental {\em in-situ} measurements.

The QMC calculation is beset with statistical fluctuations.  In
fact, the  noise of independent Monte Carlo configurations reproduces
the sample-to-sample variations of repeated experimental measurements
at the same temperature $T$ and particle number $N$ (or chemical
potential). This allows us to discuss the experimental observability
of these corrections to mean-field theory.

The Hamiltonian of $N$ interacting particles in an isotropic harmonic
trap with frequency $\omega$ is given by 
\begin{equation}
H=\sum_{i=1}^{N} \left[ \frac{p_i^2}{2m} + \frac{1}{2}m\omega^2
r_i^2\right] + \frac{1}{2}\sum_{i,j=1}^{N} V(r_{ij}), 
\end{equation}
where $V$ is the interatomic (pseudo-)potential between two particles.
A hard-core interaction with radius $a$ is customary. 

The partition function $Z$ of the system with inverse temperature
$\beta=(k_B T)^{-1}$ is given by the trace of the 
symmetrized density matrix $\rho=e^{-\beta H}$
over all states.
$Z$ satisfies the usual convolution equation:
\begin{eqnarray}
Z = \frac{1}{N!}\sum_P \int  d R \rho(R,R^P,\beta) =  \frac{1}{N!}
\sum_P \times \nonumber\\
\times \int d R  \int
 d R_2  \cdots 
\int  d R_M \rho(R,R_2,\tau)  \cdots  \rho(R_M,R^P ,\tau).
\label{convolution}
\end{eqnarray}
Here $\tau=\beta/M$, $R$ is the 3N-dimensional vector of the particle
coordinates $R=(r_1,r_2,...,r_N)$, and $R^P$ denotes the vector
with permuted labels: $R^P=(r_{P(1)},r_{P(2)},...,r_{P(N)})$ \cite{Feynman}.  
As explained elsewhere \cite{ceperley} the QMC calculation
relies on virtually exact formulas for the density
matrices $\rho(R,R',\tau)$ at the higher temperature $1/\tau$ and
performs the integral over $R, R_2, \ldots, R_M$ as well as
the sum over all permutations $P$ in Eq. (\ref{convolution}) by
Monte Carlo sampling.  Special data-handling techniques allow to
cope with very large atom numbers $N$.

As in \cite{krauth96} we consider a model system of $10,000$
particles, correponding to a critical temperature of $k_B T_c^0
\simeq 20.25 \hbar \omega$ and  a hard-core potential with radius
$a=0.0043 a_0$ ($a_0=(\hbar/ m \omega)^{1/2}$).
These values are typical for
most $^{87}$Rb experiments which are now in operation.  We
generally perform computations at different values of $\tau$,
and extrapolate to $\tau \rightarrow 0$, in
which limit the QMC formulas for the density matrices in Eq.
(\ref{convolution}) become manifestly exact.

In BEC, a single quantum state is occupied by a very large number
of bosons.  In the presence of interactions this state is
a complicated many-body wave function. As far as one particle
properties are concerned
it can, however, be described by a
single particle wave function $\psi_0(r)$
determined by the modified Gross-Pitaevskii equation
\begin{equation}
\label{n0mf}
\left(-\frac{\hbar^2\nabla^2}{2m}+\frac{m\omega^2}{2}r^2+U\,[n_0({\bf r})
+2n_T({\bf r})]-\mu_0\right)\psi_0({\bf r})=0
\end{equation}
where the densities of condensed $n_0({\bf r}) = N_0 \times|\psi_0({\bf r})|^2$
and thermally excited particles $n_T({\bf r})$ account for interactions 
between these
particles. The strength of the interaction
is given by $U=4\pi\hbar^2a/m$, where $a$ denotes again the hard sphere
radius. The factor of 2 in front of $n_T$ accounts for quantum statistical 
exchange energy.
In the semiclassical HF approximation the thermal density is given
by \cite{goldman81,giorgini97}
\begin{equation}
\label{nTHF}
n_T({\bf r}) = \frac{1}{\lambda_T^3}
g_{3/2}\left(e^{-(m\omega^2r^2/2+2U\,n({\bf r})-\mu_T)/k_BT}
\right).
\end{equation}
Here, the
thermal wavelength $\lambda_T = \hbar\sqrt{2\pi/mk_BT}$ and the
Bose function $g_{3/2}(z)=\sum_{j=1}^\infty z^j/j^{3/2}$ have been
used.  In order to obtain the density distributions of the condensate
and of the thermal component, Eqs (\ref{n0mf}) and (\ref{nTHF})
have to be solved conjointly at the same chemical potential $\mu
= \mu_0 = \mu_T$ with the constraint of the fixed particle number
$N$ (different notations for the chemical potential appear for
later convenience, cf. below).  HF neglects the collective excitations
of a more fundamental Bogoliubov theory.  However, it was observed
that these excitations do not contribute significantly to thermodynamic
properties at temperatures well above the chemical potential
\cite{dalfovo97}.  On the other hand, collective excitations are
fully included in our QMC calculation, as recently confirmed
by a comparison of two-particle correlation functions \cite{pair}.

The precise solution of the HF equation is quite difficult.  We
have obtained identical results both with an iterative procedure
\cite{dalfovo97} and an interpolation-minimization routine 
cf. \cite{krauth96}.  In a small window around the transition
temperature, the HF equations have no solution.

\begin{figure}
\centerline{ \psfig{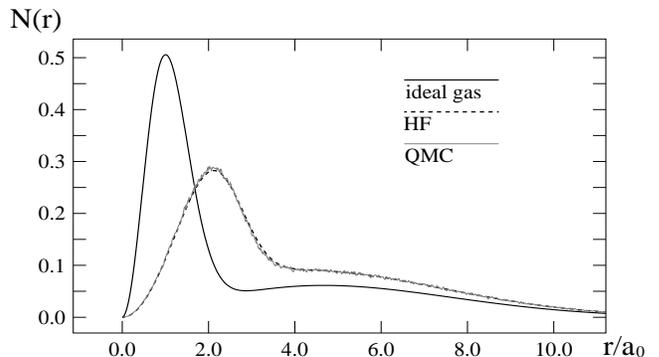} }
\caption{ Number density $N(r)$ from HF and the exact 
QMC calculations compared to the non-interacting Bose gas, both for 
$N=10,000$ particles in an isotropic trap with $\beta =
0.07/ \hbar \omega$ ($T \simeq 0.7\; T_c^0$).} 
\end{figure}

The raw output of our Monte Carlo simulations consists in histograms
of $N(r)$, the number of particles at a distance $r$ from
the center of the trap.  In Fig. 1, we compare the QMC results for
$N(r)$ with the HF solution far below the critical temperature.
We notice an excellent agreement between the two approaches.  For
comparison we also show the result for $10,000$ non-interacting
bosons at the same parameters.  The difference is very important.
In Fig. 2, we show corresponding data for a temperature closer
to the critical point.  Important qualitative information can be
obtained directly from Fig. 1 and Fig. 2 without further data analysis 
\cite{finitetau}  {\em
i)} with respect to the non-interacting solution, the number of
condensed particles (both for HF and QMC) is decreased, while their
distribution is widened.  This can be seen from the first peak in
$N(r)$, which is shifted to larger values of $r$. Notice that, even
far away from the center of the trap, the distribution of the
interacting gas is quantitatively different from the ideal gas. We
have checked that a temperature determination of these atoms within
the ideal gas model fails by about $5\%$.  {\em ii)} With respect
to the mean-field  solution, the number of condensed particles in
the exact numerical calculation is {\em in}creased.  This second
point nicely checks with recent QMC \cite{Franck} and renormalization
group calculations \cite{Stoof} for the isotropic case. These
calculations indicate that the degeneracy parameter is changed by
interaction effects beyond mean-field and that for a dilute gas
the critical temperature is increased linearly in $a/\lambda_T$.
It might be suspected that this increase is a mere consequence of
quasi-particle excitations. However, it has been shown within the
Popov approximation that quasi-particles lead rather to a minute
decrease of the critical temperature \cite{dalfovo97}.

\begin{figure}
\centerline{ \psfig{figure=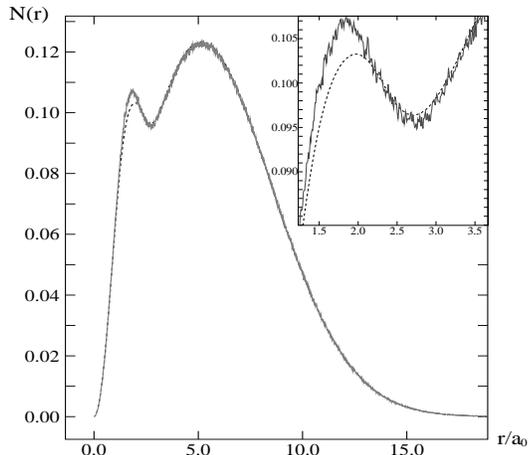,height=6cm} }
\caption{ $N(r)$ from HF and the exact 
QMC calculations for the same system as in Fig. 1, but at a higher 
temperature, $\beta=0.056/ \hbar \omega$
($T\simeq 0.88\; T_c^0$). The inset shows
the two curves in the overlap region. 
}
\end{figure}

To analyze the data we have extracted the number of condensed
particles as well as their distribution function from the QMC
calculation.  In Ref. \cite{krauth96} both were obtained from the
permutation-cycle lengths (cf. also \cite{krauth98}).  Here, however,
we have preferred to directly "fit" the QMC-generated histograms
of $N(r)$ from a very large number of samples to smooth curves
$n_0(r)$ and $n_T(r)$, as has been done to analyze experimental
{\em in situ} measurements.  We have found a nearly perfect fit to
the data, on the local density approximation level, by using
functional forms for $n_0(r)$ and $n_T(r)$ as given in Eqs
(\ref{n0mf}) and (\ref{nTHF}) with unrestricted `chemical potentials'
$\mu_0$ and $\mu_T$.  Eqs (\ref{n0mf}) and (\ref{nTHF}) are used,
but the fit procedure is {\em not} a HF approximation in disguise.
This can be seen by a simple counting argument: In the HF calculation,
the single parameter $\mu$ allows to satisfy the constraint on the
total number of particles. The HF solution, whenever it exists, usually
is
unique.  In the present case, under the same conditions,
we have one more parameter, which we fix by the condition of
minimizing the mean-square displacement between the data points
and the interpolating function.  The $\chi^2$ obtained was compatible
with purely statistical deviations for histograms containing more
than $10^7$ data points.  We note in passing that the density
$\rho(r) = N(r)/4 \pi r^2$ and the component densities are obtained
with very good precision {\em after} performing the fit of the
raw-data histogram rather than from a direct rescaling of the
data.

This fit could also provide a scheme for the accurate determination of
$T$  within experimental {\em in-situ} measurements.  In current
experiments the temperature is mostly determined by a time-of-flight
method.  Due to collisions during the expansion and the non-instantaneous
switch-off of the trap this method limits the temperature
determination to $5 \%$, so that a more accurate determination of
the critical temperature and the condensate fraction has not yet
been possible \cite{mewes96,ensher96}.

The use of the fit allows us to compute $N_0$ vs. temperature for the
set of parameters chosen. The results, extrapolated to $\tau
\rightarrow 0$, are plotted in Fig. 3 together with the HF result,
which gives a consistently smaller value of $N_0$. It should be
noticed that the HF equations, as mentioned, have no solution for
a range of temperatures close to $T_c$. In fact, on approaching
this window from below the effective chemical potential
$\mu_{\mbox{eff}}= -m \omega^2 r^2/2 - 2\; U n(r) + \mu_0 $ in Eq.
(\ref{nTHF}) approaches zero (cf. inset of Fig. 3) in the overlap
region of the two components.  Since the Bose function $g_{3/2}(z)$
diverges for $z>1$, this causes a discontinuity of the HF solution.
This problem is a finite size effect which is caused by the
nonvanishing kinetic energy of the condensate wave function.
\begin{figure}
\centerline{ \psfig{figure=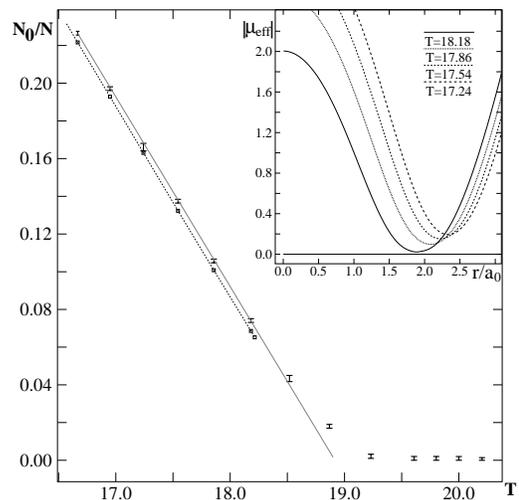,height=6.5cm} }
\caption{ Extrapolated condensate fraction $N_0/N$ from HF (lower)
and from QMC (upper) vs $T$ in units of $\hbar \omega$.  The grey
line interpolates the QMC data.  The inset illustrates the instability
of the HF solution as the effective chemical potential
$|\mu_{\mbox{eff}}(r)|$ approaches zero.
} 
\end{figure}
The correction to mean-field theory shown in Fig. 3 provides definite
support for the scenario found in the spatially homogeneous Bose
gas: Interactions lead to an increased tendency towards Bose
condensation \cite{Franck,Stoof}. There is however an important
difference of context: in a homogeneous system the density at each
point is unchanged by interactions and the shift in $T_c$ corresponds
directly to corrections to HF.  As we have seen in the present
case the trap is much more complicated since the finite-size
effects, the  mean-field itself, and finally the corrections to
mean-field all influence the density profile and the condensate
fraction. However, the finite-size effects and the mean-field itself
are already included in HF. Therefore the difference between QMC
and HF become directly comparable to the homogeneous system.  In
Ref.  \cite{Franck}, the following behavior is proposed:  $\delta
T_c/T_c^0= (T_c-T_c^0)/T_c^0 \simeq 0.34 \;a \rho^{0.34} \simeq
0.46 a/\lambda_{T_c^0}$. In our case we would expect $\delta T_c
\sim 0.065$.  This corresponds roughly to the apparently constant
offset in $T$ of the two lines in Fig. 3.

It has been  suggested \cite{houbiers97} that a direct measurement
of the density in the center of the trap $\rho(r\!=\!0)$ would be useful
to determine the deviation of the degeneracy parameter from the
mean-field result, $\rho(0) \lambda_{T_c}^3 =2.61$. Even though we
are able to extract $\rho(0)$ from QMC or from experimental data
very precisely,  the onset of condensation is not very sharp and
the finite-size effects prevent a precise determination of the
critical degeneracy parameter \cite{footscaling}.  We rather
advocate a direct comparison between HF (which is available for
finite systems) and the  data.

It is evident from an experimental point of view that the observation
of these corrections necessitates a high resolution for $T$ and
$N$.  Even for ideal experiments  the intrinsic sample-to-sample
fluctuations may prevent their observability.  Within QMC, we have
determined these fluctuations from repeated measurements of the
{\em complete} distribution of one whole sample.  To be able to
distinguish clearly the mean particle distribution from the mean-field
result, one single configuration of the $10,000$ particles is not
sufficient. We find that an average of at least $50$ independent
samples has to be taken in order to detect that there is a difference
between HF and the exact result in the sense of a Kolmogorov-Smirnov
test \cite{numericalrecipes}. For a precise evaluation (as in Fig.
3) a  much larger data pool is necessary.  Experimentally this
seems to require a nondestructive measurement scheme.

In conclusion, we are  convinced that comparison between theory
and experiment on the level treated in this paper is conceivable.
An experimental access to the mean-field corrections will be  of
outstanding theoretical interest \cite{footavail}.

Fruitful discussions with Franck Lalo{\"e} are acknowleged. 
This work was partially supported
by the EC (TMR network ERBFMRX-CT96-0002), the
Deutscher Akade\-mi\-scher Austauschdienst, 
and the Deutsche Forschungsgemeinschaft. We are grateful to the
ITP, Santa Barbara for hospitality.

%\bibliographystyle{prsty}
%\bibliography{refbec,refman}

\begin{thebibliography}{10}
\bibliographystyle{unsrt}

\bibitem{anderson95}
M.~H. Anderson, J.~R. Ensher, M.~R. Matthews, C.~E. Wieman,
 and E.~A. Cornell, Science {\bf 269},  198  (1995).


\bibitem{davis95}
K.~B. Davis, M.-O. Mewes, M.~R. Andrews, N.~J. van Druten, D.~S.
Durfee, D.~M. Kurn, and W. Ketterle, Phys. Rev. Lett. {\bf 75},
  3969  (1995).

\bibitem{bradley95} 
C.~C. Bradley, C.~A. Sackett, J.~J. Tolett, and R.~G. Hulet,
Phys. Rev. Lett. {\bf 75}, 1687 (1995);
C.~C. Bradley, C.~A. Sackett, and R.~G. Hulet,
Phys. Rev. Lett. {\bf 78}, 985 (1997).

\bibitem{bogoliubov47}
N.~N. Bogoliubov, J. Phys. (Moscow) {\bf 11}, 23 (1947).


\bibitem{gross61}
E.~P. Gross, Nuovo Cimento {\bf 20},  454  (1961);
L.~P. Pitaevskii, Sov. Phys. JETP {\bf 13},  451  (1961).

\bibitem{hutchinson97}
D.~A.~W. Hutchinson, E. Zaremba, and A. Griffin,
Phys. Rev. Lett. {\bf 78}, 1842 (1997).

\bibitem{dalfovo98}
F. Dalfovo, S. Giorgini, L. Pitaevskii, and S. Stringari,
to appear in Rev. Mod. Phys. (1998).

\bibitem{ceperley} E. L. Pollock and D. M. Ceperley, Phys. Rev.
B{\bf 30}, 2555 (1984); B{\bf 36}, 8343 (1987); D. M. Ceperley,
Rev. Mod. Phys. {\bf 67}, 1601 (1995).

\bibitem{krauth96}
W. Krauth, Phys. Rev. Lett. {\bf 77},  3695  (1996).


\bibitem{goldman81}
V.~V. Goldman, I.~F. Silvera, and A.~J. Leggett, Phys. Rev.~B {\bf 24},
2870  (1981). 

\bibitem{giorgini97}
S. Giorgini, L.~P. Pitaevskii, and S. Stringari,
Phys. Rev. ~A {\bf 54}, R4633 (1996).

\bibitem{Franck} P. Gr\"uter, D. Ceperley, and F. Lalo\"{e}, Phys.
Rev. Lett.  {\bf 79}, 3549 (1997).

\bibitem{Stoof} M. Bijlsma and H.~T.~C. Stoof, Phys. Rev.~A {\bf 54},
5085 (1996).

\bibitem{houbiers97} M. Houbiers, H.~T.~C. Stoof, and E.~A. Cornell,
Phys. Rev. ~A {\bf 56}, 2041 (1997).

\bibitem{Feynman} R.~P. Feynman, 
{\em Statistical Mechanics} (Ben\-ja\-min/Cum\-mings, Reading, MA, 1972).

\bibitem{dalfovo97}
S. Giorgini, L.~P. Pitaevskii, and S. Stringari, J.  Low Temp.
Phys. {\bf 109},  309  (1997);
F. Dalfovo, S. Giorgini, M. Guilleumas, L. Pitaevskii, and
S. Stringari, Phys. Rev. ~A {\bf 56}, 3840 (1997).

\bibitem{pair} M. Holzmann and Y. Castin (unpublished);
M. Naraschewski and R.~J. Glauber (unpublished).

\bibitem{finitetau} The QMC data shown  were obtained with 
$\tau =0.005/ \hbar \omega$.
A minute systematic shift subsists.

\bibitem{krauth98}
M. Holzmann and W. Krauth (unpublished).

\bibitem{mewes96} M.-O. Mewes, M.~R. Andrews, N.~J. van Druten,
D.~M. Kurn, D.~S. Durfee, and W. Ketterle, Phys. Rev. Lett.
{\bf 77}, 416 (1996).

\bibitem{ensher96} J.~R. Ensher, D.~S. Jin, M.~R. Matthews,
C.~E. Wieman, and E.~A. Cornell, Phys. Rev. Lett. {\bf 77}, 
4984 (1996).

\bibitem{footscaling}
The scaling behavior seems to be similar to an ideal gas, where
finite-size effects lead to a deviation of the critical degeneracy
parameter by $O(N^{-1/6})$.

\bibitem{numericalrecipes} 
 W. H. Press S. A. Teukolsky, W. T. Vetterling, B. P.
Flannery, {\sl Numerical Recipes}, 2nd edition,  Cambridge University
Press (1992).

\bibitem{footavail} The FORTRAN program for the fit procedure
and the  QMC code
are made available (from MH or
WK).
\end{thebibliography}

\end{multicols}
\end{document}